\documentclass[]{elsarticle}
\usepackage{graphicx,epsfig}
\usepackage{times}
\usepackage{graphics,bm,float}
\usepackage{amssymb,amsmath,rotate,color}

\begin{document}
\unitlength 1 cm
\newtheorem{thm}{Theorem}
\newtheorem{lem}[thm]{Lemma}
\newdefinition{rmk}{Remark}
\newproof{pf}{Proof}
\newproof{pot}{Proof of Theorem \ref{thm2}}
\newcommand{\be}{\begin{equation}}
\newcommand{\ee}{\end{equation}}
\newcommand{\bearr}{\begin{eqnarray}}
\newcommand{\eearr}{\end{eqnarray}}
\newcommand{\nn}{\nonumber}
\newcommand{\vk}{\vec k}
\newcommand{\vp}{\vec p}
\newcommand{\vq}{\vec q}
\newcommand{\vkp}{\vec {k'}}
\newcommand{\vpp}{\vec {p'}}
\newcommand{\vqp}{\vec {q'}}
\newcommand{\bk}{{\bf k}}
\newcommand{\bp}{{\bf p}}
\newcommand{\bq}{{\bf q}}
\newcommand{\br}{{\bf r}}
\newcommand{\up}{\uparrow}
\newcommand{\down}{\downarrow}
\newcommand{\fns}{\footnotesize}
\newcommand{\ns}{\normalsize}
\newcommand{\cdag}{c^{\dagger}}

\definecolor{red}{rgb}{1.0,0.0,0.0}
\definecolor{green}{rgb}{0.0,1.0,0.0}
\definecolor{blue}{rgb}{0.0,0.0,1.0}

\title{Effects of correlations on honeycomb lattice in ionic-Hubbard Model }
\author[IAUM]{M. Ebrahimkhas\corref{cor}}
\author[IAUM]{Z. Drezhegrighash}
\author[IAUM]{E. Soltani}
\address[IAUM]{Department of physics, Mahabad Branch, Islamic Azad university, Mahabad 59135, Iran}
\cortext[cor]{Corresponding author}

\begin{abstract}

In a honeycomb lattice  the symmetry has been broken by adding an ionic potential and a single-particle gap was generated in the spectrum.
We have employed the iterative perturbation theory (IPT) in dynamical mean field approximation method to study
 the effects of competition between $U$ and $\Delta$ on energy gap and
renormalized Fermi velocity. We found, the competition between the single-particle gap parameter and the Hubbard potential closed
the energy gap and restored the semi-metallic phase, then the gap is opened again in Mott insulator phase. For a fixed $\Delta$  by 
increasing $U$, the renormalized Fermi velocity $\tilde{v_F}$ is decreased, but change in $\Delta$, for a fixed $U$,
has no effects on  $\tilde{v_F}$. The difference in filling factor is calculated for various
number of $U, ~\Delta$. The results of this study can be
implicated  for gapped graphene e.g. hydrogenated graphene.

\end{abstract}


\maketitle

\section{Introduction}

Many mechanisms can open an energy gap in metallic or semi-metallic state, such as: Mott-insulator transition or
charge and spin density wave on nested Fermi surface~\cite{Urig}. In addition,
 competition between two interactions can close and suppress the energy gap~\cite{Garg1}. For studying the opening
 and suppression of energy gap in spectrum of a honeycomb lattice, we started with a
simple tight binding Hamiltonian, which the substrate can induce a symmetry breaking through an
 ionic potential of strength $\Delta$ and adding on-site repulsive interaction $U$~\cite{Garg2}.
 
 The ground state of the honeycomb lattice when $\Delta\neq0, ~U=0$ is a band insulator on the strongly correlated limit~\cite{Sorella-Tosatti},~\cite{Watanabe}.
 This immediately leads to a band gap of magnitude $2\Delta$, with site occupancies $n_{lower~band}=2,~n_{higher~band}=0$.
This gap decreases with increasing $U$, and no-double occupancy constraint imposed by large $U$~\cite{EbrahimkhasIH}. In opposite limit, $i.e.$
 $\Delta \ll U$ the system can be described in terms of an effective massive Dirac theory.
 The intermediate phase can be found 
 between two insulator and massive Dirac fermions, which is obeyed massless Dirac theory. The first purpose of this paper is 
tackle to these questions:
 What would be the result of competition between $U$ and $\Delta$ on occupancy of each sub-lattice in honeycomb lattice? 
 Or what would be the effects of  $U$ and $\Delta$ on the energy gap? 
 Another target of this paper is  studying the effects of  competition between $U$ and $\Delta$ on renormalization of
 Fermi velocity.
 
  There is a widespread consensus on the large potential of graphene for electronic applications~\cite{Novoselov},~\cite{Ebrahimkhas-cjp}. The low electronic density of states near the Fermi energy and zero band-gap at neutrality point in graphene  exhibit a small ON/OFF switching ratio, due to this problem   the application of graphene for charge based logic devices are inhibited ~\cite{Britnel}. It
has been shown, that pure graphene exhibits weak anti ferromagnetic properties  at near room temperature~\cite{Sorella}. The carbon-based systems such as graphene are  the most attractive objects for hydrogen storage~\cite{Yazyev}.
The hydrogen adsorption on graphene is an interesting idea for two reasons: first: a band gap is induced~\cite{Balog} and second: the hydrogenated
graphene sheet is converted to a hydrogen storage device~\cite{Boukhvalov}.
The results of this paper could be generalized to hydrogenated graphene.

In semi-metal-insulator transition (SMIT) unlike to metal insulator transition (MIT), there is no Kondo resonance 
corresponding to quasi-particle at the Fermi level~\cite{Fritz}. Therefore, SMIT can be described by renormalized Fermi velocity $\tilde{v_F}$
 instead of spectral weight of such resonant state. The Fermi velocity is an order parameter which is characterized a Dirac liquid state~\cite{Jafari2009}.
We use dynamical mean field theory (DMFT) for probing  the effects of $U$ and $\Delta$  within paramagnetic phase
 in ionic-Hubbard model.
The DMFT is exact in limit of infinite coordination numbers~\cite{Georges}, but for lower dimensions the local self-energy (here is k-independent)
becomes only an approximate description~\cite{Wunsch}. Therefore the critical values of some parameters in DMFT approximation on honeycomb
lattice maybe overestimated~\cite{EbrahimkhasDMFT}. But we should note the overall picture of output numerical results is expected to hold.

 The Phase separation in 2D  Hubbard model was studied with the dynamical cluster approximation (DCA)~\cite{Jarrell}.  In  the DMFT method the interactions in a lattice are
  mapped to an impurity problem which is embedded self-consistently in a host.  Therefore the DMFT  neglects spatial correlations, But in DCA we assume that correlations are
  short range.  The original lattice is mapped to a periodic cluster of specific size, which is embedded in a self-consistent host. Therefore the  correlations in same  range with
   the  cluster size are  considered accurately, while the other interactions with longer length than cluster are described at the mean-field approximation. 
   In limit of $\Delta \rightarrow 0$, the ionic-Hubbard model and Hubbard model are similar, when we use DCA and DMFT in this limit ($\Delta \rightarrow 0$),
   we found different results on the same model and same lattice.  These differences are in magnitude of critical $U_c$ and border of the phase
    regions,    but   the overall descriptions of phase diagram and phase separation are expected to hold.
   The metal-insulator transition  has obtained by  cluster DMFT (CDMFT), is different from that of the single-site DMFT. In CDMFT with cluster size larger
    than  2,  the quasi particle weight  is k-dependent and nonzero, but in the single-site DMFT, the quasi particle weight  is k-independent and vanishes
     continuously at the MIT region~\cite{Imada}.
   In Ref.~\cite{Kocharian}, the variational cluster approximation (VCA) is applied to calculate local electron correlations in bipartite square and honeycomb lattices in Hubbard
    model, they found, in honeycomb lattice electron density displayed smooth metal–insulator transition with  continuous evolution. The square lattice experienced 
   metal-insulator transition, but  the electron density  in square lattice displayed discontinuity with spontaneous transition~\cite{Kocharian}.
   The phase transition  in the ionic Hubbard model has been  investigated in a two-dimensional square lattice by determinant quantum Monte  Carlo (DQMC).
   The competition between staggered potential and on site potential lead to the  phase transition from Mott insulator to Metallic  and band insulator~\cite{Scalettar1}.

  The ionic-Hubbard (IH) model  has been studied in 1-D and 2D~\cite{Manmana}.  The DMFT approximation has been employed to study of such model~\cite{Nourafkan},
  this technique is implemented to study the phase transitions and phase diagrams of honeycomb lattice in IH model~\cite{EbrahimkhasIH} and a square lattice was studied by determinant quantum Monte Carlo method~\cite{Scalettar2}.
The results can be related to physics of graphene and hydrogenated graphene for specific magnitude of $\Delta$ and $U$.
 
\section{Model and Method}

The ionic-Hubbard model (IHM) on the honeycomb lattice is described by this Hamiltonian,

\bearr
   H=-t\sum_{i\in A,j\in B, \sigma} 
   \left( c^\dagger_{i\sigma}c_{j\sigma}+{\rm h.c.}\right)-\mu\sum_{i}n_i  \nn \\ 
   +\Delta\sum_{i\in A}{n_i}-\Delta\sum_{i\in B}{n_i}+U\sum_j n_{j\down} n_{j\up},
   \label{IHH}
\eearr

where $t$ is the nearest neighbour hopping, $\Delta$ is a staggered one-body potential that alternates sign between site in sub-lattice
$A$ or $B$ and $U$ is the Hubbard repulsion. The chemical potential is $\mu=U/2$ at half-filling and so the average occupancy is
$\frac{<n_A>+<n_B>}{2}=1$. This model  represented a band insulator with energy gap $2\Delta$ at non-interacting limit,
$U=0$.
In opposite limit, $U\gg\Delta$, the system is in Mott insulator state. What's the intermediate phase? 
The semi-metallic character restored and massive Dirac fermions will become massless,  
as a result of increasing the Hubbard potential, $U$. 
It has been demonstrated that by DMFT method can describe and understand  band insulator (BI)-semi-metal (SM)-Mott insulator (MI) transition~\cite{Garg1}. The first step is
introducing interaction Green's function in bipartite lattice,

\be
G(\vk, \omega^+) =
\left( {\begin{array}{cc}
 \zeta_A(\vk, \omega^+) & -\epsilon(\vk)  \\
 -\epsilon(\vk) & \zeta_B(\vk,\omega^+)  \\
 \end{array} } \right)
 \label{Matrix-G}
\ee

where $\vec{k}$ is the momentum vector in first Brillouin zone, $\epsilon(\vec{k})$ is the energy dispersion for the honeycomb lattice,
and $\zeta_{A(B)}=\omega^+\mp\Delta+\mu-\Sigma_{A(B)}(\omega^+)$, with $\omega^+=\omega+i0^+$. 
The local Green's function corresponding to each sub-lattice can be written as,

\begin{equation}
   G_{\alpha}(\omega^+)=\sum_{\vec{k}}G_{\alpha~\alpha}(\vec{k},\omega^+)=\zeta_{\bar{\alpha}}(\omega^+)
   \int^{\infty}_{-\infty}d\epsilon \frac{\rho_{0}(\epsilon)}{\zeta_{A}(\omega^+)\zeta_{B}(\omega^+)-\epsilon^2}
   \label{GFint}
\end{equation}

where $\alpha=A(B)$, $\bar{\alpha}=B(A)$ are corresponding to each sub-lattice,
and $\rho_{0}(\epsilon)$ is the bare DOS of the honeycomb lattice (graphene).
The details of calculations are done in previous work~\cite{EbrahimkhasIH}. The DOS of an interaction system can be calculated by,

\begin{equation}
 \rho_{\alpha}(\omega)=\Sigma_{\vec{k}}\Im~Tr[G_{\alpha}(\vec{k},\omega^+)]/\pi. 
\label{DOS}
\end{equation} 
 
According to particle-hole symmetry at half-filling in honeycomb lattice, we know $\rho_{A}(\omega)=\rho_{B}(-\omega)$.
 The total DOS eventually obtained via $\rho(\omega)=\rho_{A}(\omega)+\rho_{B}(\omega)$. 
 The DOS is necessary for obtaining the energy gap for each pair of $U,\Delta$.
In computation of energy gap for each pair of $U,\Delta$, the density of state is essential and slope of $\rho(\epsilon)$ near Dirac points determined renormalized  Fermi velocity in SM phase~\cite{Jafari2009}.

\section{Results and discussion}

\begin{figure}[tb]
\begin{center}
\includegraphics[width=10cm,height=6cm,angle=0]{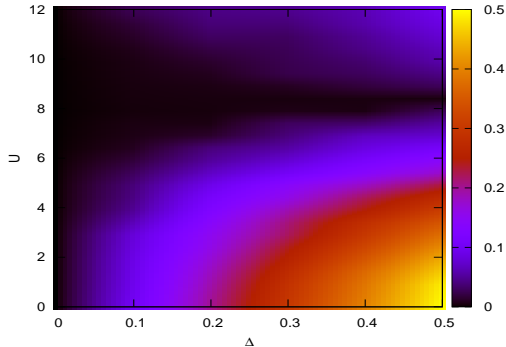}
\vspace{-3mm}
\caption{(Color online) The density plot of the energy gap in $U,~\Delta$ plate. The band insulator and Mott insulator are
separated by semi-metal phase. }
\label{Egap.fig}
\end{center}
\vspace{-3mm}
\end{figure}

 The filling factors of each sub-lattice, $n_{A}, n_{B}$ and energy gap have been calculated by DMFT outputs.
 In Fig.~\ref{Egap.fig}, the density plot of the energy gap is plotted $vs$ , $U,~\Delta$. In calculation of $E_{gap}$, the $U,~\Delta$ have been changed in $0.1t$ steps, then for better resolution we used interpolation method to obtain continuous density plot.  We can observe , the density plot of energy gap is
 in excellent agreement to Fig. 4 of Ref.~\cite{EbrahimkhasIH}. The energy gap vanishes in the semi-metallic phase. In this phase diagram one can
 find graphene in $U/t\sim 3.0-4.0$ and $\Delta/t\sim 0.04-0.05$, where the system still remains in semi-metallic phase~\cite{Wehling}. 
 So the graphene can be SM, despite a symmetry breaking ionic potential of strength $\Delta \sim 110-150~meV$ ($t\sim 2.7eV$). The renormalization of the gap magnitude is not the only reason of the Hubbard correlations U . It also influences by other spectral features such as the life-time of quasi particles, e.g. in hydrogenated graphene~\cite{Harberer}.
 In this DMFT approximation, results may have overestimated the upper bound, $U_c$ ~\cite{Wu}. For improving this calculations, we can use cluster-DMFT~\cite{Wu}. 
 It is expected the upper bound of $U$ push to down, but our estimate of $\Delta$  is not expected to change much. The phase diagram for a  square lattice has obtained
  with other  interesting details in Refs.~\cite{Scalettar1},~\cite{Scalettar2} by DQMC method. In square lattice we could find three phase region: band and Mott insulator and metallic.

 At low energy regions the dispersion becomes $\epsilon_{\vec{k}}=\pm v_{F} k$ and the DOS have
  linear energy dependence~\cite{NetoRMP},
  
  \begin{equation}
   \rho(\epsilon)=2\pi v_{F}^{-2} \mid\epsilon\mid
   \label{rho}
\end{equation}
 
 where $v_{F}$ is the bare Fermi velocity at Fermi point. In non-interacting limit, by adding an ionic potential $\Delta$, the
 energy gap is created and the energy gap is the order of $2\Delta$. Therefore the DOS at low energy obtained as,
 
  \begin{equation}
   \rho(\epsilon)=2\pi v_{F}^{-2} \sqrt{\epsilon^2 +\Delta^2}
   \label{rhodelta}
\end{equation}

\begin{figure}[tb]
\begin{center}
\includegraphics[width=10cm,height=6cm,angle=0]{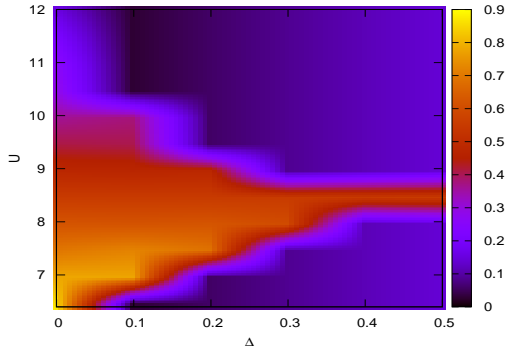}
\vspace{-1mm}
\caption{(Color online) The density plot of Fermi velocity is plotted. The calculation of $\tilde{v_F}$ is valid in SM phase.
The Hubbard term can affect on  $\tilde{v_F}$ but ionic potential has no effect on it.  }
\label{vf.fig}
\end{center}
\vspace{-1mm}
\end{figure}

In Fig.~\ref{vf.fig}, we have shown the density plot of renormalized Fermi velocity ,$\tilde{v_{F}}$, in $U-\Delta$ plane. As can
be seen for a fixed $\Delta$, by increasing $U$, Fermi velocity decreases. This results is adapted with previous works~\cite{Jafari2009}. In this figure we used interpolation method, so the results are acceptable only in region with SM phase.
In a fixed $U$, when $\Delta$ changes we can see Fermi velocity remains with no-variation.
As $U$ increases the slope of DOS increases and spectral weight is transferred to higher energies.
In semi-metallic phase: the Fermi velocity has inverse relation with $\sqrt {\partial\rho(\epsilon) / \partial \epsilon}$, so we see reduction in $v_F$ by
increasing $U$.
The DOS around the Fermi level, at energy scales above $\Delta$ has $\vee$ shape. By increasing $\Delta$ the slope of this '$\vee$' shape
doesn't change. 

 Therefore at SM phase at constant $U$ by increasing $\Delta$,
the slope of DOS and  magnitude of $v_F$ will be fixed, consequentially.
In BI and MI phase near Dirac point one can find an opening gap which is expanded by increasing ionic potential.

\begin{figure}[tb]
\begin{center}
\includegraphics[width=6cm,height=5cm,angle=0]{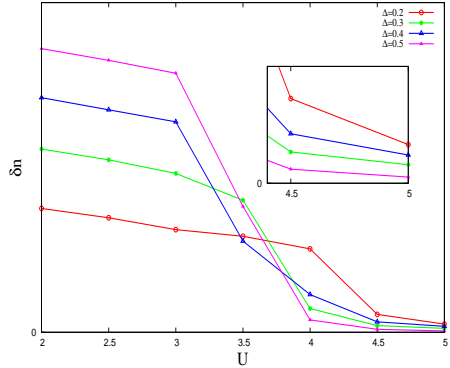}
\vspace{-3mm}
\caption{(Color online) The difference in filling factor for $\Delta/t=0.2,0.3,0.4,0.5$ $vs$ Hubbard potential is shown.  }
\label{dn.fig}
\end{center}
\vspace{-3mm}
\end{figure}

The difference in filling factor $\delta n=(n_B - n_A)/2$  is plotted in Fig.~\ref{dn.fig}. The $\delta n$ is calculated for different ionic potential
as a function of $U$. In band insulator phase for a $(U=0,~\Delta\neq0)$ the difference in filling factor is 1 and 
$\langle n_B \rangle=2,~\langle n_A \rangle=0$. When $U$ is increased, the occupation of lower band will be depleted slowly. 
In extremity of $U$ both band have same occupation number and  $\delta n=0$. In calculation of $\delta n$ for various ionic
potential $vs$ Hubbard potential, we observed: the $\delta n$ has dropped in $3 \leqslant U \leqslant 4$ interval for $\Delta/t=0.4,~0.5$. 
The variation in $\delta n$ for $\Delta/t=0.2$ is smaller in compared with other values of $\Delta$. In the inset of  Fig.~\ref{dn.fig}, we have zoomed in the $4 \leqslant U \leqslant 5$ region. We found, for small ionic
potential $e.g.$ $\Delta=0.2t$, the $\delta n$ has been vanished in  $U\gg5t$. The results of Fig.~\ref{dn.fig} is in good agreement with out put of
 Fig.~\ref{Egap.fig}.
 
 \section{Summary and discussions}
 
 We have studied  the influences of correlation in ionic-Hubbard model on the honeycomb lattice by IPT-DMFT, and we have calculated the energy gap $vs$  $U,~\Delta$.
 For $\Delta=0$ the system is in semi-metallic phase and the energy gap is zero. The renormalized Fermi velocity is decreased
  by increasing $U$. In $\Delta\neq0$ region,  for $U<U_{c1}$ the system is in a band insulator phase, when $U$ is increased
  the energy gap will be closed. For $U>U_{c2}$ the energy gap was opened again and the Mott insulator phase was appeared . In SM phase, the increasing
  of $U$ can decrease the Fermi velocity of quasi-particle near Fermi level, But in semi-metallic phase the increasing of $\delta$ doesn't
  change the $\tilde{v_F}$. 
  The difference in occupation number can shows the phase transition of the system. The smooth shape of $\delta n$ is according to results of~\cite{Kocharian}.
   Our calculations  demonstrated the $\Delta n$ for 
  large $\Delta$  has  fast change in $3 <U/t<4$ interval, for small energy gap and $\Delta$ the occupation of each site has  smooth variation. 
  The overall picture of the phase transition in honeycomb lattice is same as other results on bipartite honeycomb and square lattice with CDMFT, VCA and DQMC method. Conceptually our results showed, the Hubbard model at half filling exhibits similar behaviour in the square and honeycomb 2D structures. In limit of $\Delta \rightarrow 0$, when we compare CDMFT  and DMFT(IPT) results,  we found different results on same lattices. These differences are not in overall picture and phase separation predictions but are in $U_c$ magnitude and in borders of phase diagram. A reason is in technical details of methods; In CDMFT, the local electron correlations are considered and the self-energy has $\vec{k}$ dependency, but in our calculations we didn't consider these interactions, so By neglecting and renormalization of some interactions we found these differentiations.


\begin{thebibliography}{29}

\bibitem{Urig} Michał Karski, Carsten Raas, Götz S. Uhrig, Phys. Rev. B {\bf 77}, 075116 (2008).
\bibitem{Garg1} Arti Garg, H. R. Krishnamurthy, and Mohit Randeria, Phys. Rev. Lett. {\bf 97}, 046403 (2006).
\bibitem{Garg2} Arti Garg, H. R. Krishnamurthy, and Mohit Randeria,
Phys. Rev. Lett. {\bf 112}, 106406 (2014).
\bibitem{Sorella-Tosatti} N. Gidopoulos S. Sorella, E. Tosatti, Eur. Phys. J. B, {\bf 14}, 217 (2000).
\bibitem{Watanabe} T. Watanabe, S. Ishihara, J. Phys. Soc. Jpn. {\bf 82}, 034704 (2013).
\bibitem{EbrahimkhasIH} M. Ebrahimkhas, S. A. Jafari, Eur. Phys. Lett. {\bf 98} 27009 (2012).
\bibitem{Novoselov} K. S. Novoselov, A. K. Geim, S. V. Morozov, D. Jiang, Y. Zhang, S. V. Dubonos, 
I. V. Grigorieva, and A. A. Firsov, Science {\bf 306}, 666 (2004);
K. S. Novoselov, A. K. Geim, S. V. Morozov, D. Jiang, 
M. I. Katsnelson, I. V. Grigorieva, S. V. Dubonos, A. A. Firsov, Nature {\bf 438}, 197 (2005); A. Bostwick, T. Ohta, T. Seyller, K. Horn, and E. Rotenberg, Nature Physics {\bf 3}, 36 (2007); A. H. Castro Neto, F. Guinea, N. M. R. Peres, K. S. Novoselov, A. K. Geim, Rev. Mod. Phys. {\bf 81}, 109 (2009).
\bibitem{Ebrahimkhas-cjp} M. Ebrahimkhas, Chin. J. of Phys, {\bf 52}, 1602 (2014).
\bibitem{Britnel} L. Britnell, R. V. Gorbachev, R. Jalil, B. D. Belle,
F. Schedin, A. Mishchenko, T. Georgiou, M. I. Katsnel-
son, L. Eaves, S. V. Morozov, N. M. R. Peres, J. Leist,
A. K. Geim, K. S. Novoselov, and L. A. Ponomarenko, Science
{\bf  335}, 947 (2012).
\bibitem{Sorella} S. Sorella and E. Tosatti, EPL {\bf 19 }, 699 (2007).
\bibitem{Yazyev} O. V. Yazyev and L. Helm, Phys. Rev. B {\bf 75}, 125408 (2007).
\bibitem{Balog} R. Balog, B. Jørgensen, L. Nilsson, M. Andersen, E. Rienks, M. Bianchi, M. Fanetti, E. Lægsgaard, A. Baraldi, S. Lizzit, Z. Sljivancanin, F. Besenbacher, B. Hammer, T. G. Pedersen, P. Hofmann, and L. Hornekær, Nature Mat. {\bf 9}, 315 (2010).
\bibitem{Boukhvalov} D. W. Boukhvalov, M. I. Katsnelson, and A. I. Lichtenstein, Phys. Rev. B {\bf 77}, 035427 (2008).
\bibitem{Fritz} L. Fritz, M. Vojta, Rep. Prog. Phys. {\bf 76}, 032501 (2013).
\bibitem{Georges}  A. Georges, et al, Rev. Mod. Phys. {\bf 68}, 13 (1996);
M. Caffarel, W. Krauth, Phys. Rev. Lett. {\bf 72} , 1545 (1994).
\bibitem{Wunsch} B. Wunsch, F. Guinea, F. Sols, New J. Phys. {\bf 10},  103027  (2008).
\bibitem{EbrahimkhasDMFT} M. Ebrahimkhas, Phys. Lett. A {\bf 375}, 3223 (2011).
\bibitem{Jarrell} A. Macridin, M. Jarrell, Th. Maier, Phys. Rev. B {\bf 74} 085104 (2006).
\bibitem{Imada} Y. Z. Zhang, M. Imada, Phys.Rev.B {\bf 76} 045108 (2007). 
\bibitem{Kocharian} A. N. Kocharian, Kun Fang, G.W. Fernando, A.V. Balatsky, Journal of Magnetism and Magnetic Materials {\bf 10}, 007 (2014).
\bibitem{Scalettar1}  K. Bouadim, N. Paris, F. Hebert, G. G. Batrouni, and R. T. Scalettar, Phys. Rev. B {\bf 76}, 085112  (2007).
\bibitem{Manmana}  S. R. Manmana, V. Meden, R. M. Noack, K. Schoenhammer, Phys. Rev. B {\bf 70}, 155115 (2004); K. Byczuk, M. Sekania, W. Hofstetter, A. P. Kampf, Phys. Rev. B {\bf 79}, 121103(R) (2009); M. Hafez Torbati, Nils A. Drescher, Götz S. Uhrig,  Phys. Rev. B {\bf 89}, 245126 (2014); F. Mancini, Tech. Rep., University of Salerno (2005).
Jafari2009
\bibitem{Nourafkan} R. Nourafkan, G. Kotliar, Phys. Rev. B {\bf 88}, 155121 (2013).
\bibitem{Scalettar2} N. Paris, K. Bouadim, F. Hebert, G. G. Batrouni, and R. T. Scalettar, Phys. Rev. Lett. {\bf 98}, 046403 (2007)
\bibitem{Jafari2009} S. A. Jafari, Eur. Phys. Jour. B {\bf 68}, 537 (2009).
\bibitem{Wehling} T. O. Wehling, E. Sasioglu, C. Friedrich, A. I. Lichtenstein, M. I. Katsnelson, S. Blugel, Phys. Rev. Lett., {\bf 106}, 236805 (2011).
\bibitem{Wu} W. Wu, Yao-Hua Chen, Hong-Shuai Tao, Ning-Hua Tong, and
Wu-Ming Liu, Phys. Rev. B {\bf 82}, 245102 (2010).
\bibitem{Harberer} D. Haberer, et. al. Physica Status Solidi (b), {\bf 248}, 2639 (2011).
\bibitem{NetoRMP} A. H. Castro Neto, F. Guinea, N. M. R. Peres, K. S. Novoselvo, A. K. Geim, Rev. Mod. Phys. {\bf 81}, 109 (2009).



\end{thebibliography}
\end{document}